\documentstyle[preprint,aps,epsfig]{revtex}
\tightenlines
\newcommand{\vsig}{\mbox{\boldmath $\sigma$ \unboldmath}}
\newcommand{\veps}{\mbox{\boldmath $\epsilon$ \unboldmath}}

\begin{document}

\title{\bf $\eta^\prime$ photoproduction near threshold }

\author{Qiang Zhao\thanks{Present address: Dept. of Physics, Univ. of Surrey, 
Guildford, GU2 7XH, UK; 
Email address: qiang.zhao@surrey.ac.uk.}}
\address{ Institut de Physique Nucl\'eaire, F-91406 Orsay Cedex, France } 

\date{\today}

\maketitle  
  
\begin{abstract}
In this work, the $\eta^\prime$ meson photoproduction
near threshold is studied in the quark model framework.
A pseudovector effective Lagrangian is introduced for the $\eta^\prime NN$
coupling and the newly published data from the SAPHIR Collaboration 
provide good constraints to this parameter. 
Corrections of order $O(1/m_q^2)$ for the electromagnetic interaction
vertex are taken into account, which produce corrections 
of order $O(1/m_q^3)$ to the transition amplitude 
for $\gamma p\to \eta^\prime p$. 
Some low-lying resonances, $S_{11}(1535)$, $P_{13}(1720)$, 
and $P_{13}(1900)$ are found to have significant 
contributions. A bump structure around 
$E_\gamma\approx$ 2 GeV is found arising from the $n=3$ terms in the 
harmonic oscillator basis.
The beam polarization asymmetries are predicted and can be tested 
against the forthcoming data from GRAAL.

\end{abstract}
\vskip 1.cm

PACS numbers: 12.39.-x, 13.60.-r, 14.20.G, 14.40.Gs

\newpage

\section{Introduction}

The study of $\eta^\prime$ meson photoproduction becomes more 
and more attractive in both experiment and theory. 
However, 
within a long period, only few data 
were available from ABBHHM~\cite{ABBHHM68} and AHHM~\cite{AHHM}. 
According to those results, it was difficult to even establish 
the feature of the resonance excitations. 
This situation challenged theoretical attempts to study 
$\eta^\prime$ photoproduction, $\gamma p\to \eta^\prime p$. 
In Ref.~\cite{RPI-eta-prime-95}, a hadronic 
model was proposed with a single resonance 
$P_{11}(2050)$ excited in this reaction, which thus accounted for the
possible strong peak near threshold. 
However, the roles of those low-lying resonances 
in this reaction have not been explained.
A more self-consistent attempt was made by Li ~\cite{li-eta-prime}
through a quark model approach 
based on the assumption that the $\eta^\prime$ meson couples to the
constituent quark {\it via} the same pseudovector coupling  
as that of the $\eta$. 
Although the available data 
at that time did not permit detailed investigation of this channel, 
the author showed that a strong peak near threshold could be produced 
by the low-lying resonances, especially dominated by the $S_{11}(1535)$. 
Meanwhile, a small bump 
around $E_\gamma=2$ GeV, which comes from the $n=3$ resonance
excitations, was predicted. However, in that work, two resonances, 
$P_{13}(1900)$ and $F_{15}(2000)$, which are close to the $\eta^\prime$
production threshold, had not been included as little 
was known about them from Ref.~\cite{PDG98}. we are left with 
questions about the role played by these two resonances 
and more interestingly, whether the 
$\gamma p\to \eta^\prime p$ can 
provide signals for the existences of these two resonances. 
This is related to the motivation of
searching for ``missing resonances" in the meson photoproductions. 
To answer such questions, more precise measurements for this reaction 
are needed.

The recently published data from the SAPHIR Collaboration~\cite{bonn98} 
brought 
the possibility of further systematical study of the 
resonance excitations in $\gamma p\to \eta^\prime p$. 
With the photon energy covering a range from threshold 
(1.45 GeV) to 2.6 GeV for the first time, 
the resonance excitations were established experimentally 
in $\eta^\prime$ meson photoproduction.
The steep rise and fall of the total cross section near threshold
indicated the dominance of resonance excitations. 
In Ref.~\cite{bonn98}, the resonance structure was explained as
due to two resonance excitations ($S_{11}(2090)$ and $P_{11}(2100)$). 
Such a prescription 
led to the arbitary assumption that low-lying resonance effects could be 
completely neglected, and risked overlooking 
influences of small cross sections to the angular distributions.
As an example, we recall that in the study of $\eta$ meson
photo- and electro-production, the small contribution from the $D_{13}(1520)$ 
produces significant effects which result in the deviation of the 
angular distributions from the dominant $S$-wave in the low energies.
Moreover, resonances other than the $S_{11}(2090)$ and 
$P_{11}(2100)$ have masses very close to the $\eta^\prime$ production 
threshold, such as the $P_{13}(1900)$. 
Therefore, one should also include their threshold
effects in this reaction.
Based on the above considerations, we believe that a self-consistent 
treatment to systematically include all the resonances in the 
$\gamma p\to \eta^\prime p$ reaction is required, and the quark model provides
an ideal starting point~\cite{li97}.

In this work, we follow the same scheme as in Ref.~\cite{li-eta-prime}
to study the $\eta^\prime$ photoproduction at tree level in the quark model.
There are a number of new features in this work:
1) The electromagnetic interaction is expanded
to $O(1/m_q^2)$, which will partly account for the relativistic 
corrections beyond the NRCQM. 
2) Two resonances, $P_{13}(1900)$ and $F_{15}(2000)$, which are  
attributed to representation $[{\bf 70}, {\bf ^2 8}, 2, 2, J]$, 
have been included. 
These two resonances are not well determined experimentally. 
In $\gamma p\to \eta^\prime p$, we expect that some signals 
for these two resonances can be clarified.
3) In this study, the equal velocity frame (EVF) 
is adopted for the Lorentz boost~\cite{fostor82}. 
As shown in Ref.~\cite{zhao-eta}, the EVF can 
boost the particle momentum close to a realistic value.
An {\it ad hoc} form factor is avoided in the EVF for the spatial 
integrals.
4) The new data have 
 better contraints to the 
nucleon pole terms, that sheds light on the  
$\eta^\prime NN$ coupling constant from the quark model. 
5) In addition to the study of the differential and total cross sections, 
the beam polarization asymmetry is predicted, which can be tested by the 
forthcoming data from GRAAL and JLab in the near future.

In Section II, a brief introduction to this model is given. 
Results and analysis are presented in Section III. 
Conclusions are drawn in Section IV. 

\section{The model}

The $\eta^\prime NN$ interaction is introduced at quark level 
through the effective Lagrangian for the quark-meson vertex~\cite{li97}:
\begin{equation}\label{lagrangian}
L_{eff}=\sum_{j}\frac{1}{f_{\eta^\prime}}\overline{\psi}_j\gamma_\mu^j
\gamma_5^j\psi_j\partial^\mu\phi_{\eta^\prime} \ ,
\end{equation}
where $\psi_j$ ($\overline{\psi}_j$) represents 
the $j$th quark (anti-quark) field in the nucleon, and $\phi_{\eta^\prime}$
is the meson field. 
It is still not elementarily clear if the $\eta^\prime$ couples to the 
nucleon through a pseudoscalar or pseudovector coupling, or even both.
However, as pointed out in Ref.~\cite{yaouanc},  
the operators for the pseudoscalar and pseudovector coupling
 have the same leading order expression at quark level.

We take the assumption that the $\eta^\prime NN$ coupling
satisfies the Goldberger-Treiman relation, 
which means that we can relate 
the quark-meson coupling to the 
$\eta^\prime NN$ coupling through
\begin{equation}
g_{\eta^\prime NN}=\frac{g_A M_N}{f_{\eta^\prime}} \ .
\end{equation}
The coupling $g_{\eta^\prime NN}$
is treated as a parameter that will be determined by the experimental data.

The electromagnetic interaction for the three-quark baryon system 
can be expanded to order $O(1/m_q^2)$: 
\begin{equation}
\label{EM}
H_{em}=(\frac{\omega_\gamma^3}{2})^{\frac 12}
\sum_{j} (h_j^c + h_j^i) e^{i{\bf k}\cdot {\bf r}_j} \ ,
\end{equation}
where 
\begin{eqnarray}
\label{cm}
h_j^c & \equiv & [ e_j {\bf R} \cdot\veps 
-\frac{e_j}{2m_j}\vsig\cdot (\veps\times \hat{\bf k}) \nonumber\\
& & +\frac{1}{4M_T}(\frac{e_j}{m_j}-\frac{e_T}{M_T})
\vsig\cdot(\veps\times{\bf P}) ] \ , 
\end{eqnarray}
and 
\begin{eqnarray}
\label{internal}
h_j^i &\equiv & e_j({\bf r}_j-{\bf R})\cdot \veps \nonumber\\
& & +\frac 14(\frac{e_j}{m_j}-\frac{e_T}{M_T})\vsig\cdot
[\veps\times (\frac{{\bf p}_j}{m_j}-\frac{{\bf P}}{M_T})] \ ,
\end{eqnarray}
where $\hat{\bf k}$ is the unit vector of the photon momentum,
$\hat{\bf k}\equiv {\bf k}/\omega_\gamma$; 
$h_j^c$ and $h_j^i$ 
are operators for the c.m. and internal motions 
of the three-quark baryon system,
respectively. 
The charge and mass of the $j$th quark are denoted as $e_j$ and $m_j$, 
respectively, while $e_T$ and $M_T$ 
denote the total charge and mass of the baryon, respectively. 
The position and momentum for the 
$j$th quark are ${\bf r}_j$ and ${\bf p}_j$, respectively, 
while ${\bf R}$ and ${\bf P}$ are for the c.m. motion 
of the baryon system.

We explicitly write the $S_{11}(1535)$ transition amplitude given 
by Eq.~\ref{EM} as follows:
\begin{eqnarray}
\label{S11-transverse}
{\cal M}_{S_{11}}&=&
\frac{2M_{S_{11}} e^{-\frac{{\bf k}^2+{\bf q}^2}{6\alpha^2}} }
{(s-M_{S_{11}}^2+i M_{S_{11}}\Gamma_{S_{11}})}
\left[ \frac{\omega_\eta^\prime}{\mu_q}-(\frac{\omega_\eta^\prime}
{E_f+M_N}+1)
\frac{2{\bf q}^2}{3\alpha^2} \right] \nonumber\\
&\times & \left\{ \frac{1}{6} (\omega_\gamma+\frac{{\bf k}^2}{2m_q})
\vsig\cdot\veps_\gamma \right. \nonumber\\
& & + \left. \frac{\omega_\gamma}{9 m_q}(\mu_p-\mu_0)|{\bf k}| |{\bf q}| \cos (\theta)
\vsig\cdot\veps_\gamma \right. \nonumber\\
& & + \frac{\omega_\gamma}{36 m_q}(\mu_p-\mu_0)
\vsig\cdot\left({\bf q}
\times({\bf k}\times \veps_\gamma )\right)\nonumber\\ 
& & - \left. \frac{\omega_\gamma}{36 m_q}(\mu_p-\mu_0)\vsig\cdot{\bf k}
{\bf q}\cdot\veps_\gamma \right\} \ ,
\end{eqnarray}
where $\omega_\eta^\prime$ denotes the meson energy in the 
$\eta^\prime N$ c.m. system, and $\mu_q$ is the reduced mass of two quarks which 
equals $m_q /2 $ here. The first term in the curly bracket is the 
formula with the electromagnetic interaction expanded to order $O(1/m_q)$, i.e. 
from the first lines in Eq.~\ref{cm} and ~\ref{internal}. 
This term is the leading 
contribution which has been extensively used 
in literature~\cite{li-eta-prime,li97}. 
Taking into account the meson interaction vertex, which is also expanded 
to order $O(1/m_q)$, the transition amplitudes of the first term in the curly 
bracket of Eq.~\ref{S11-transverse} has been rigorously expanded 
to order $O(1/m_q^2)$. In other words, the  
electromagnetic interaction of order $O(1/m_q^2)$ 
will introduce corrections at order $O(1/m_q^3)$
to the transition amplitudes.
These corrections
are given by the terms proportional to $(\mu_p-\mu_0)$ which is the 
anomalous magnetic moment of the proton at leading order. Here,
$\mu_p$ and $\mu_0$ are the proton magnetic moment and the nuclear magneton, 
respectively, and they are defined as
\begin{equation}
\langle N_f| \sum_j \frac{e_j}{2m_j} \vsig_j |N_i\rangle 
\equiv \langle N_f|\mu_p \vsig |N_i \rangle \ ,
\end{equation}
and
\begin{equation}
\mu_0 \equiv \frac{e_T}{2M_T} .
\end{equation}
Note that in Eq.~\ref{S11-transverse}, the proton charge, $e_T$, has been 
moved out of the transition amplitude.

The formalism for the $\eta^\prime$ production 
is the same as for the $\eta$ apart from the inclusion
of order $O(1/m_q^2)$ corrections to the electromagnetic interaction. 
We even adopt the quark model 
parameters in the $\eta$ production here, i.e. the harmonic oscillator 
strength $\alpha=384.5$ MeV and $m_q=330$ MeV. The main decay channels
of the $S_{11}(1535)$ are $\pi N$ and $\eta N$, 
for which the branching ratios 0.45 and 
0.55 are adopted, respectively. These values are the 
same as derived in the $\eta$ meson production~\cite{zhao-eta}. 
It should be pointed out that at high energies the 
phase space will permit small branching ratio for a off-shell 
resonance decaying into higher mass channels. 
This feature will change the on-shell branching ratio values 
accordingly. However, at leading order, the influences from 
such deviations are negligibly small. Within a wide energy regions, 
the energy-dependence of the total decay 
width is still dominated by decay of the lower mass channels.
Therefore, although the total width of the $S_{11}(1535)$ 
is occupied by the $\pi N$ and $\eta N$, 
the off-shell $S_{11}$ will 
permit small branching ratio to $\eta^\prime N$ 
channel. 

\section{Results and analysis}

The newly published photoproduction data from SAPHIR 
provide for the first time 
the near-threshold angular distributions for 
$\gamma p \to \eta^\prime p$.
Although the uncertainties are still large, 
the resonance excitations can be recognized clearly. 

We present calculations of the angular distributions 
in Fig.~\ref{fig:(1)} for seven energies and compare with the 
data of Ref.~\cite{bonn98}. 
Near threshold (Fig.~\ref{fig:(1)}a), the cross section is dominated 
by the $S$-wave. 
The quite flattened angular distribution 
cannot be explained by the {\it t}-channel
meson exchanges, which are generally forward peaked.
In Fig.~\ref{fig:(1)}a,
the solid curve denotes the calculation with the middle energy 
$E_\gamma=1.490$ GeV. 
It shows that the threshold cross section is dominated by the $S_{11}(1535)$
excitation. 
We illustrate the $S_{11}(1535)$ 
influence by switching off its contribution. 
Comparing the dot-dashed curve to the solid one, we see that
the $S_{11}(1535)$ strongly enhances the cross section near threshold.
To show the possible uncertainties arising 
from the photon energy, we also present the calculations with $E_\gamma$ 
deviating 20 MeV from the middle values, i.e. $E_\gamma=1.470$ (dashed curve)
 and 1.510 GeV (dotted curve).

The $S_{11}$ excitation also accounts for the large cross section 
in the energy region up to $E_\gamma \approx 1.8$ GeV. 
In Fig.~\ref{fig:(1)}b and c, the forward peaking is found to be 
produced by the 
interferences between the $S$-wave and $P$-wave amplitudes. 
Without the $S_{11}$, the angular distribution exhibits only
a weak forward peak (dot-dashed curve in Fig.~\ref{fig:(1)}b).
It should be noted that 
the excitation threshold of the $P_{13}(1900)$
is very close to the 
$\eta^\prime$ production threshold. However, the threshold 
energy is not ideal for identifying the $P_{13}(1900)$ signal due to 
the strong $S_{11}(1535)$ contribution. At $E_\gamma=1.490$ GeV, 
the effects of the $P_{13}(1900)$ absence cannot be seen clearly. 
But in Fig.~\ref{fig:(1)}c, we show that the $P_{13}(1900)$ is
important to account for
the forward peaking at $E_\gamma=1.690$ GeV. The dashed curve 
in Fig.~\ref{fig:(1)}c 
denotes the calculations without the $P_{13}(1900)$, 
which completely changes the angular distributions at forward angles. 
The difference between the solid and dashed curves, 
are found from the interferences between the $S$ and $P$-wave 
amplitudes (dominant ones), which result in the forward peaking. 
This feature might suggest that the forward peaking 
can serve as a signal of the 
$P_{13}(1900)$ excitation in $\gamma p \to \eta^\prime p$.

In the energy region, $1.74 < E_\gamma < 2.04$ GeV, the angular 
distributions show some structures which might come from 
resonance excitations. We find that apart from 
the $S_{11}(1535)$, $P_{13}(1720)$ and $P_{13}(1900)$, 
contributions from $n=3$ terms become important. 
Although the calculation cannot 
account for the obvious increase at 90$^\circ$ in Fig.~\ref{fig:(1)}d, 
one can see that the cross section turns to level off
at this energy. 
Qualitatively, such a structure favors 
a smaller $P$-wave strength. In Fig.~\ref{fig:(1)}d, 
the dashed curve denotes the 
angular distribution without the $P_{13}(1900)$.

The dashed curve in Fig.~\ref{fig:(1)}e denotes the calculations without 
$n=3$ terms. The backward peak comes from the $P_{13}(1900)$ contribution. 
At this energy region, the $P_{13}(1900)$ cross section
has dropped down significantly. However, its influence still 
makes sense as the dotted curve in Fig.~\ref{fig:(1)}e illustrates that 
without the $P_{13}(1900)$, the dashed curve is significantly changed. 
Comparing the three curves, 
we learn that the $P_{13}(1900)$ and the
$n=3$ terms play important roles in producing the
forward peak, and simultaneously attenuate the backward peaking tendency.
The same feature continues up to 2.14 GeV. As shown in Fig.~\ref{fig:(1)}f, 
without the $P_{13}(1900)$ and $n=3$ terms (dotted curve) 
the forward peak cannot be reproduced.

Up to 2.44 GeV photon energies, 
the resonance excitations of the low-lying resonances with $n\le 2$ 
will be competing with the excitations from higher harmonic oscillator shells. 
It becomes very complicated to distinguish
 the individual resonance excitations because of their small cross sections
 and wide mass overlaps.
Here, the energy evolution turns out to be important for such a collective 
description. Impressively, in Fig.~\ref{fig:(1)}g, we find that 
the forward peaking character
keeps in the degenerate $n=3$ terms, while the energy evolution
of the cross sections are also perfectly satisfied. 
It should be noted that the backward peak in Fig.~\ref{fig:(1)}g is produced 
by the {\it u}-channel processes.

In Fig.~\ref{fig:(2)}, the total cross sections are presented
 in comparison with the 
old data from Ref.~\cite{ABBHHM68,AHHM}, and the new data 
from SAPHIR~\cite{bonn98}. 
Good agreement with the data is obtained over large energy regions. 
The solid curve shows the full calculation of our model. 
Explicitly, the role of the strong $S_{11}(1535)$ 
can be shown by comparing the 
results of the exclusive $S_{11}$ excitation (dashed curve), and the 
calculations without the $S_{11}$ (dot-dashed curve).
It shows that the sharp peak near threshold is produced by the 
$S_{11}(1535)$ excitation. 

Interestingly, a bump 
is {\it automatically} produced around $E_\gamma \approx 2$ GeV. 
This structure is 
a unique signal of the $n=3$ resonance contributions in the harmonic 
oscillator basis. As shown by the dotted curve, 
the second bump does not show up without $n=3$ excitation.
The descrepancy between the solid and dotted curve 
suggests that the interferences between the low-lying resonance excitations
and the $n=3$ terms play a key role in producing the second bump
around $E_\gamma=2$ GeV.
In the data from SAPHIR, 
a very similar structure shows up at the same energy. 
But taking into account the large errors, we need more precise 
data to justify it.

In Fig.~\ref{fig:(2)}, 
large partial cancellations exist between the resonance  
excitations and the nucleon pole terms near threshold. 
With the $S_{11}$ excitation and the nucleon pole terms, 
the cross sections are over-estimated significantly in the lower 
energy region $E_\gamma < 2.0$ GeV. We do not show this result in order 
to keep the figure clear to read. 
Such an enhancement, however, 
is cancelled by the resonance terms. 
Removing the $S_{11}$ contribution, and comparing the results 
{\it with} (dot-dashed curve) and {\it without} (heavy dotted curve) 
the nucleon pole terms, 
we see that the nucleon pole terms cancel the amplitudes significantly
over a large energy range. 
Interestingly, such a cancellation 
is sensitive to the $\eta^\prime NN$ coupling and
we find that very little freedom is left for the nucleon pole terms.
When a coupling $\alpha_{\eta^\prime}=0.22$ is adopted, 
the $\eta^\prime NN$ coupling $g_{\eta^\prime NN}=1.66$ can be 
derived.

Among those low-lying resonances of $n\le 2$, 
only the $P_{13}(1900)$ and $F_{15}(2000)$ of 
representation $[{\bf 70}, {\bf ^2 8}, 2, 2, J]$ are above the 
threshold of the $\eta^\prime$ production. 
In general, 
for a resonance $N^*$ above the $\eta^\prime$ threshold,
one can relate its helicity amplitudes 
to its exclusive total cross section 
at its mass position $M_R$: 
\begin{equation}
\label{A-12}
\sigma_{tot}(\gamma p\to N^* \to \eta^\prime p)= 
\frac{M_N}{M_R}\frac{b_{\eta^\prime}}{\Gamma_R}  
2 \{ |A_{\frac 12}|^2+ |A_{\frac 32}|^2 \} \ ,
\end{equation}
where $b_{\eta^\prime}$ is the branching ratio of the $N^*$ 
decay into $\eta^\prime N$ channel. 
$A_{\frac 12}$ and $A_{\frac 32}$ are the two independent photon excitation
helicity amplitudes of the resonance. 
This relation is model-independent, and thus
given sufficient information for the photon interaction vertex, 
one would in principle be able to derive information 
about the meson interaction vertex
for the resonance. 
However, at present the status of these two resonances, 
$P_{13}(1900)$ and $F_{15}(2000)$, has not 
been well-established. Information about their decay modes, 
branching ratios, as well as the photon excitation 
helicity amplitudes, is not available. To go as far as possible
based on the present situation, we shall use the quark model 
calculations as input to study the excitations of these two resonances
in the $\eta^\prime$ production. 
On the one hand, we calculate the exclusive total cross section 
$\sigma_{tot}(\gamma p\to N^* \to \eta^\prime p)$ 
in this model. From Eq.~\ref{A-12}, 
we can derive the quantity 
$\xi\equiv\{ |A_{\frac 12}|^2+ |A_{\frac 32}|^2 \}^{\frac 12}$
 for three values of the banching ratio $b_{\eta^\prime}=0.1$, 0.2 and 0.3.
The sensitivity of $\xi$ to the branching ratio $b_{\eta^\prime}$
can be shown. 
On the other hand, we separately calculate the helicity amplitudes
$A_{\frac 12}$, $A_{\frac 32}$ for $\xi$ 
in the NRCQM. 
Assuming the NRCQM provides the leading order calculations 
for the resonance photo-excitations, we can determine the
magnitudes of the $\eta^\prime N$ branching ratios 
for these two resonances by comparing the results of the two methods.

In Table~\ref{tab:(1)}, the quantities derived through the two methods
are shown. 
Several lessons can be learned here: 
i) For both resonances,
their branching ratios to $\eta^\prime N$ 
might be as large as 20\%.
ii) The $P_{13}(1900)$ becomes very interesting due to the feature
that it favors a larger branching ratio to $\eta^\prime N$ channel, 
and its mass is very close to the threshold of $\eta^\prime$ production. 
iii) The NRCQM calculations of the helicity amplitudes 
give the same order of magnitude for $\xi$, which might suggest 
that the meson coupling has been reasonably estimated. 

We do not expect the results in Table~\ref{tab:(1)} 
to be precise 
due to the present lack of data, as well as shortcomings 
of this approach. 
However, the prediction of the order of magnitude of the 
$\eta^\prime NN$ coupling
can be regarded as reasonable, although so far 
it has not been well-determined.
The $\eta^\prime$ meson production 
might be a possible channel to determine the 
$P_{13}(1900)$ and $F_{15}(2000)$ experimentally.

We extend the calculation to the beam polarization asymmetry $\Sigma$. 
In terms of the helicity amplitudes, this has the form
\begin{equation}
\Sigma=-\mbox{Re} \{ H_1(\theta) H_4^*(\theta)-H_2(\theta) H_3^*(\theta)\} \ ,
\end{equation}
where $H_{1,2,3,4}$ are the four independent helicity transition 
amplitudes that have been extensively discussed 
in the literature (See, for example Ref.~\cite{fasano92}). 
The angle $\theta$ is given by ${\bf k}\cdot {\bf q}
=|{\bf k}| |{\bf q}| \cos(\theta)$ in the c.m. system of the 
final state hadrons.
Normalized by the differential cross section, 
the beam polarization asymmetry is presented in Fig.~\ref{fig:(3)}a for three
energies, $E_\gamma=1.65$ (solid), 1.6 (dashed) and 1.5 GeV (dotted).
Near threshold, the asymmetries are found to be small. 
It shows that the $\Sigma$ is sensitive to the energy. 
Large negative asymmetries  
are produced at $\theta\approx 140^\circ$. 
We find the nodal structure is governed by the $S$ and $P$-wave 
interferences. 
To be more clear, we study the $S$ and $P$-wave 
interference effects at $E_\gamma=1.65$ GeV
in Fig.~\ref{fig:(3)}b by eliminating the $P_{13}(1900)$ 
to produce the dashed curve. Comparing the dashed curve to the 
solid one, it shows that the $P_{13}(1900)$ strongly shifts the positive node
to the backward angles, and enhances it significantly. 
Then by removing the $P_{13}(1720)$, we obtain the dotted curve 
in Fig.~\ref{fig:(3)}b, which exhibits small asymmetries and is governed by the 
dominant $S$-wave amplitudes. 
It should be noted that the influence
of the $D_{13}(1520)$ becomes very weak in the $\eta^\prime$ production. 
Qualitatively, 
a stronger $D$-wave will attenuate the negative trend and enhance 
the positive one. 
We expect that data from the GRAAL Collaboration 
and JLab will provide tests of this prediction.

\section{Conclusions}

In summary, we studied $\eta^\prime$ meson photoproduction 
near threshold by introducing an effective 
Lagrangian for the quark meson vertex in the quark model. 
The {\it s}-channel resonances are systematically taken into account. 
With the data from SAPHIR, the resonance excitations near threshold 
are established in this reaction. Contrary to the explanation 
of Ref.~\cite{bonn98}, we find that
the $S_{11}(1535)$ accounts for the 
steep peaking near threshold. 
The $S$ and $P$-wave interference is found to be important in 
reproducing the angular distributions near threshold. This feature 
might be useful for identifying the $P_{13}(1900)$ signal in this channel.
In this model, a bump structure is {\it automatically} produced 
due to the contributions of $n=3$ terms in the harmonic oscillator 
basis. A similar structure appears in the data but needs 
to be justified with more precise measurement. 
The nucleon pole terms can be reasonably constrained in this approach 
and provide an estimate of the $\eta^\prime NN$ coupling of 
$g_{\eta^\prime NN}=1.66$. 

Expanding the photon interaction to $O(1/m_q^2)$, we obtain  
corrections to order $O(1/m_q^3)$ for $\gamma p\to \eta^\prime p$.
We find that near threshold such corrections do not change the calculations 
significantly in comparison with calculations at $O(1/m_q^2)$.
This suggests that the leading order contributions 
based on the NRCQM contain the main characters of such a 
non-perturbative process. Such an empirical starting point 
is still helpful for us to understand the 
nucleon resonance internal structures.

\acknowledgements

The author thanks F. Klein for providing the data of the SAPHIR Collaboration
and useful discussions. 
Very fruitful comments from Z.-P. Li and B. Saghai are gratefully acknowledged.
We also wish to thank J.-P. Didelez, E. Hourany and M. Guidal for 
their interests in this work. Special thanks go to J. Al-Khalili 
for help on the improvement of the language.
Financial support from IPN-Orsay is acknowledged.

\vspace*{1cm}

\begin{table}
\caption{ The dependence of the quantity $\xi$ on the $\eta^\prime N$ 
branching ratio in this model for the $P_{13}(1900)$ and 
$F_{15}(2000)$. Independent calculations in the NRCQM are presented 
as a comparison. }
\protect\label{tab:(1)}
\begin{center}
\begin{tabular}{l|c|c}
 & $P_{13}(1900)$ & $F_{15}(2000)$ \\[1ex]\hline
$M_R$ (MeV) & 1900 & 2000 \\[1ex]
$\Gamma_R$ (MeV) & 400 & 400 \\[1ex]
$\xi_1(b_{\eta^\prime}=0.1)$ ( GeV$^{\frac 12}$)
& 5.10$\times 10^{-3}$ & 2.64$\times 10^{-3}$\\[1ex]
$\xi_2(b_{\eta^\prime}=0.2)$ ( GeV$^{\frac 12}$)
& 3.61$\times 10^{-3}$ & 1.86$\times 10^{-3}$\\[1ex]
$\xi_3(b_{\eta^\prime}=0.3)$ ( GeV$^{\frac 12}$)
& 2.62$\times 10^{-3}$ & 1.52$\times 10^{-3}$\\[1ex]\hline
$\xi_{\gamma p\to N^*}$ ( GeV$^{\frac 12}$)
& 2.84$\times 10^{-3}$ & 1.74$\times 10^{-3}$\\[1ex]
$A_{\frac 12}$ ( GeV$^{\frac 12}$)
& $-2.74\times 10^{-3}$ & $1.04\times 10^{-3}$ \\[1ex]
$A_{\frac 32}$ ( GeV$^{\frac 12}$)
& $0.732\times 10^{-3}$ & $-1.39\times 10^{-3}$ \\[1ex]
\end{tabular}
\end{center}
\end{table}    


\begin{figure}
\caption{ Angular distributions for 7 energy bins are compared 
to the data from SAPHIR~\protect\cite{bonn98}. 
The calculations at the middle energies 
of each bins are presented by the solid curves. See text for the
notations for the dashed, dotted and dot-dashed curves. }
\protect\label{fig:(1)}
\end{figure}
\begin{figure}
\caption{ Total cross sections compared with the 
data~\protect\cite{bonn98,ABBHHM68,AHHM}. 
The solid curve denotes the full calculation 
of this model. The dashed curve denotes the exclusive contribution 
from the $S_{11}(1535)$ excitation, while the dot-dashed denotes full 
calculation with the $S_{11}(1535)$ switched off. 
The dotted curve denotes calculation 
without $n=3$ terms, while the heavy-dotted curve, without 
the $S_{11}(1535)$ and the nucleon pole terms.}
\protect\label{fig:(2)}
\end{figure}
\begin{figure}
\caption{ Beam polarization asymmetries predicted by this model. 
In (a), asymmetries for three energies $E_\gamma=1.65$ (solid), 
1.6 (dashed) and 1.5 GeV (dotted), are shown. 
In (b), the solid curve is the same as that in (a). 
The dashed curve denotes calculation without the $P_{13}(1900)$, 
while the dotted curve without both $P_{13}(1900)$ and $P_{13}(1720)$. }
\protect\label{fig:(3)}
\end{figure}

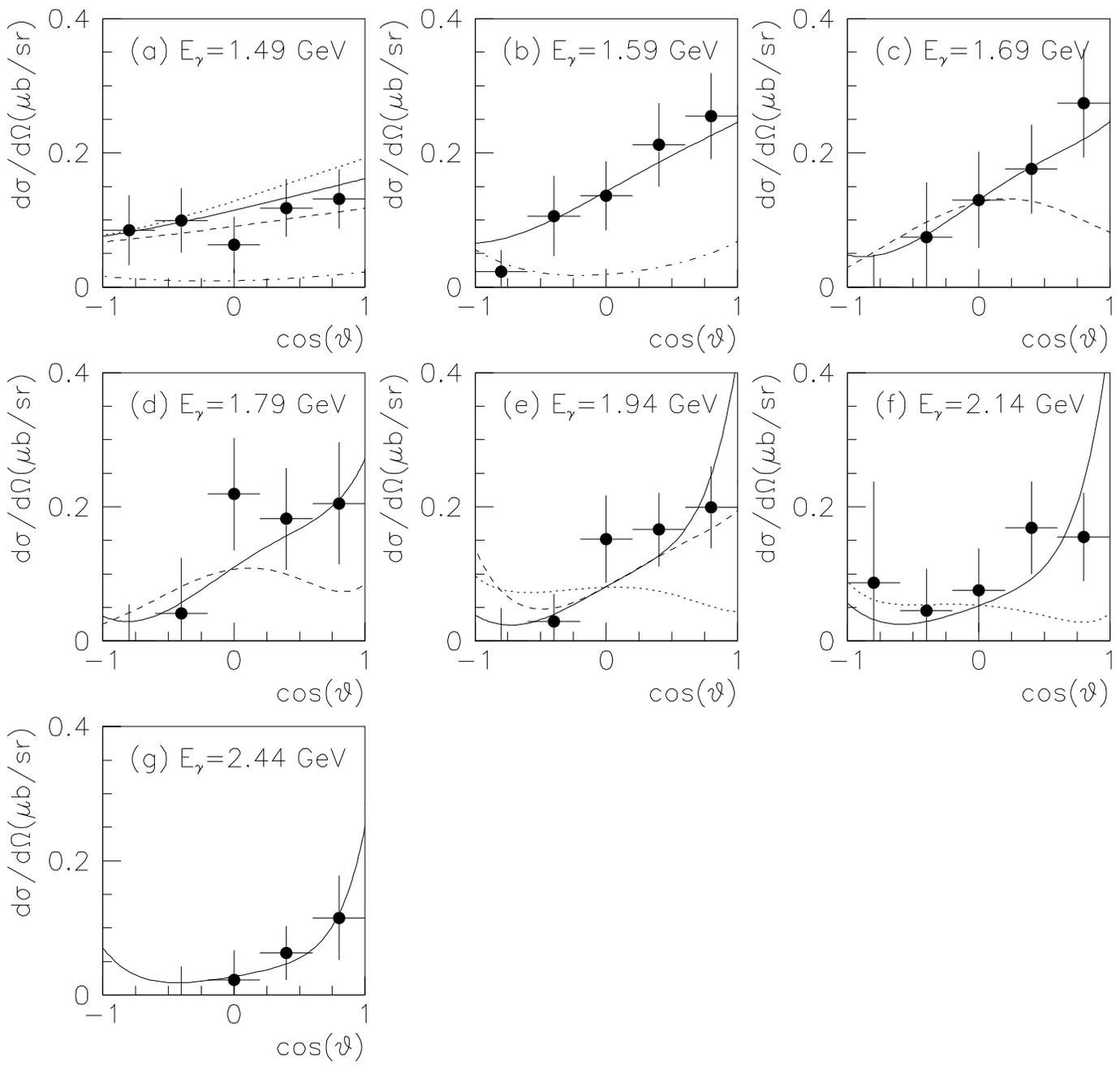
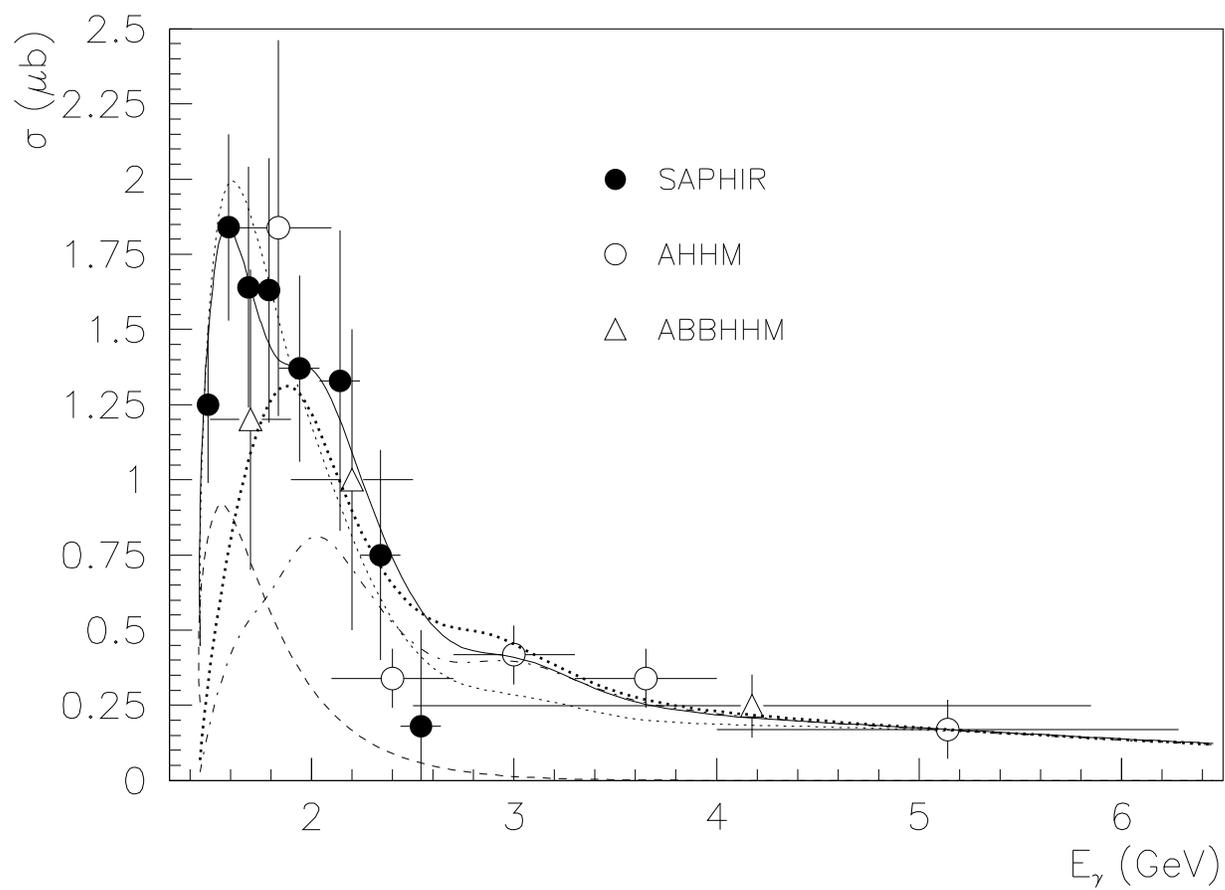
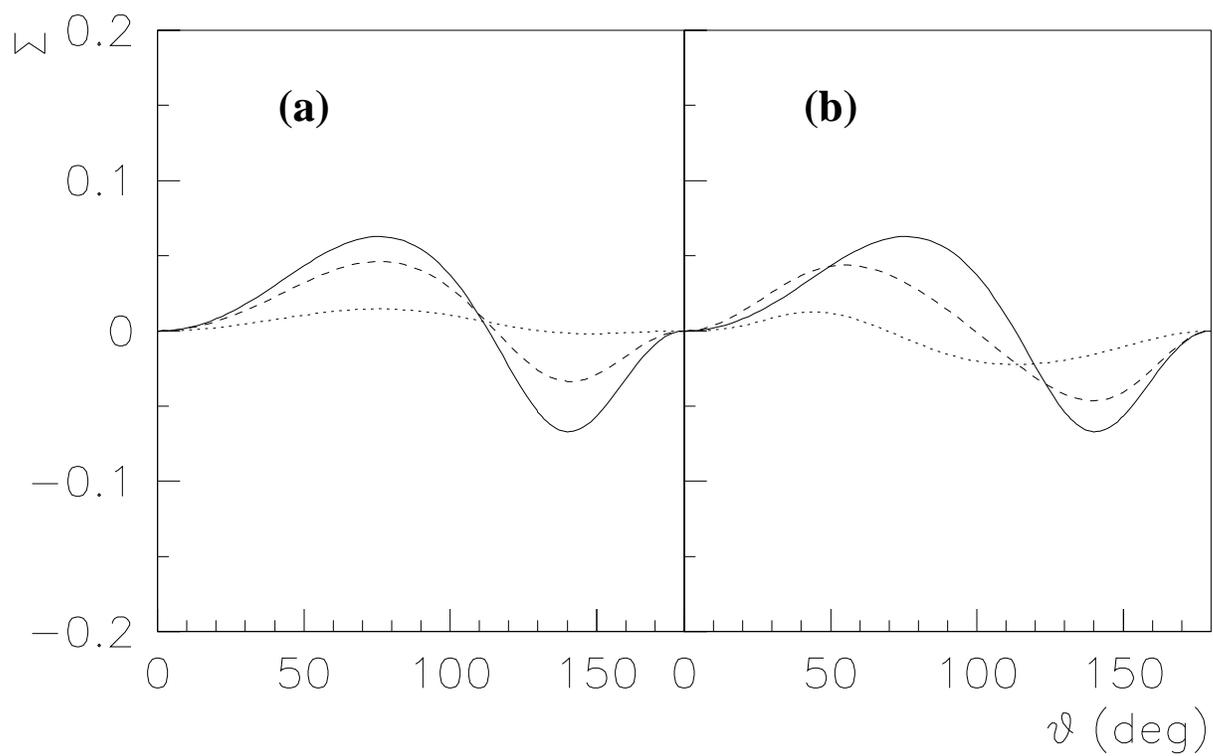
\end{document}